\documentclass[aps,prd,nofootinbib,superscriptaddress]{revtex4-2}

\usepackage[T1]{fontenc}
\usepackage{amsmath,amssymb,mathtools}
\usepackage{bm}
\usepackage{booktabs}
\usepackage{microtype}
\usepackage{array}
\usepackage{xcolor}
\usepackage{hyperref}

\begin{document}
\raggedbottom

\title{Parity-violating spatially covariant gravity at total derivative order $d=5$}

\author{Jin-Jian Song}
\affiliation{School of Physics, Sun Yat-sen University, Guangzhou 510275, China}

\author{Xian Gao}
\email[Corresponding author: ]{gaoxian@mail.sysu.edu.cn}
\affiliation{School of Physics, Sun Yat-sen University, Guangzhou 510275, China}
\affiliation{Guangdong Provincial Key Laboratory of Quantum Metrology and Sensing,\\ Sun Yat-sen University, Zhuhai 519082, China}
	
\date{July 22, 2026}

\begin{abstract}
We extend the polynomial construction of parity-violating spatially covariant gravity (SCG) to total derivative order $d=5$, where $d=d_{\mathrm{t}}+d_{\mathrm{s}}$ counts the total number of temporal and spatial derivatives.  After organizing the monomials by $(d_{\mathrm{t}},d_{\mathrm{s}})$ and reducing them using integrations by parts, tensor symmetries, three-dimensional curvature identities, the Schouten identity, and the Cayley-Hamilton relation, we obtain a $59$-element basis: $2$, $40$, and $17$ monomials in the sectors $(0,5)$, $(2,3)$, and $(4,1)$, respectively.  A direct inspection separates $35$ monomials containing neither the lapse velocity $\mathcal{L}_{\bm{u}}\ln N$ nor $\mathcal{L}_{\bm{u}}K_{ij}$ from $11$ lapse-velocity monomials and $13$ monomials containing higher normal derivatives of the spatial metric.  The latter two sectors require a dedicated degeneracy analysis.  For tensor perturbations about a spatially flat cosmological background, the acceleration-free part of the manifestly first-order-in-time sector contains $16$ basis elements, whose quadratic action depends on only four combinations of coefficients.  These combinations generate helicity-odd corrections proportional to $k/a$ and $(k/a)^3$ in the kinetic and gradient functions.  We derive two relations that enforce luminal phase velocity for both circular polarizations while still allowing helicity-dependent kinetic normalization and damping.  

\end{abstract}

\maketitle


\section{Introduction}
\label{sec:intro}

The observed acceleration of the Universe and the unresolved nature of the dark sector motivate systematic tests of gravity beyond general relativity.  Scalar-tensor theories provide a broad and flexible framework for such modifications, while higher-derivative interactions enlarge the space of possible dynamics and effective-field-theory corrections.  Their construction must, however, control the additional degrees of freedom that generically accompany higher temporal derivatives.  Considerable progress has been made by exploiting degeneracy conditions and unitary-gauge formulations, which allow higher-derivative scalar-tensor interactions to be organized while avoiding, or at least isolating, the unwanted Ostrogradsky mode \citep{Horndeski:1974wa,Deffayet:2011gz,Kobayashi:2011nu,Zumalacarregui:2013pma,Langlois:2015cwa,Langlois:2017mxy,DeFelice:2018ewo,DeFelice:2021hps}.

Parity violation provides a particularly sharp probe of gravitational dynamics beyond general relativity because it distinguishes the two circular tensor polarizations.  Representative models include four-dimensional Chern-Simons modified gravity and its scalar-tensor generalizations \cite{Chern:1974ft,Deser:1981wh,Jackiw:2003pm,Alexander:2009tp,Crisostomi:2017ugk}, as well as parity-odd sectors of Ho\v{r}ava gravity \cite{Horava:2008ih,Horava:2009uw,Takahashi:2009wc,Pereira-Dias:2011rva,Wang:2012fi,Gong:2021jgg}.  Parity-odd modifications have also been investigated in connection-based formulations in which torsion or nonmetricity carries the additional geometric structure.  Representative examples include Nieh-Yan-modified teleparallel models \cite{Li:2020xjt,Li:2021wij,Cai:2021uup}, parity-violating symmetric-teleparallel and scalar-nonmetricity theories \cite{Li:2021mdp,Li:2022vtn,Chen:2022wtz,Conroy:2019ibo,Yu:2025job}, and parity-odd metric-affine or Palatini constructions \cite{Iosifidis:2020dck,Sulantay:2022sag,Iosifidis:2021bad}.  The geometric origin of the Nieh-Yan density and related torsion-induced parity-odd terms was discussed in \cite{Nieh:1981ww,Chatzistavrakidis:2020wum}.  Dedicated analyses have also examined the stability and cosmological tensor phenomenology of broader teleparallel parity-violating models \cite{Li:2022mti,Hohmann:2022wrk,Rao:2023doc}.

The direct detection of gravitational waves and the development of planned space-based observatories provide increasingly sensitive opportunities to test such effects \cite{LIGOScientific:2016aoc,LIGOScientific:2017vwq,LISA:2017pwj}.  In gravitational-wave propagation, parity-odd interactions can generate helicity-dependent damping and phase velocities, conventionally described as amplitude and velocity birefringence \cite{Alexander:2017jmt,Zhao:2019xmm,Gao:2019liu,Jenks:2023pmk}.  Existing observations have already constrained such effects using GW170817 and compact-binary catalogs, while stochastic-background searches and future detector networks provide complementary probes \cite{Nishizawa:2018srh,Okounkova:2021xjv,Callister:2023tws,Zhang:2025kcw,Wang:2020cub,Wang:2021gqm,Wu:2021ndf,Califano:2023aji,Seto:2020zxw}.  During inflation they may also produce chiral primordial gravitational waves and nonvanishing TB/EB correlations in the cosmic microwave background \cite{Satoh:2007gn,Satoh:2010ep,Creminelli:2014wna,Cannone:2015rra,Lue:1998mq,Saito:2007kt,Contaldi:2008yz,Gluscevic:2010vv,Sorbo:2011rz,Moretti:2024fzb,Qiao:2019hkz,Li:2024fxy}.  Precision measurements of CMB polarization and analyses of cosmic birefringence provide complementary motivation for searches for parity-odd signatures in cosmology \cite{Minami:2020odp,Eskilt:2022cff,Komatsu:2022nvu}.  Higher-spatial-derivative parity-odd terms are natural from an effective-field-theory viewpoint and can generate characteristic momentum dependence in tensor propagation \cite{Gong:2021jgg,Zhu:2022dfq,Zhu:2022uoq,Zhu:2023rrx}.  
Scalar-induced gravitational waves have also been studied extensively in parity-violating gravity theories \cite{Zhang:2022xmm,Feng:2023veu,Zhang:2023scq,Zhang:2024vfw,Feng:2024yic,Feng:2026pcs}.

A direct generally covariant construction of parity-violating scalar-tensor interactions was developed in \cite{Crisostomi:2017ugk}.  There, the Levi-Civita tensor is combined with the Riemann tensor and covariant derivatives of a scalar field, up to second order, to enumerate parity-odd couplings extending Chern-Simons gravity.  This method works directly with covariant scalar-tensor building blocks and makes contact with familiar geometric interactions.  In that construction, the combinations that are healthy in unitary gauge are selected by imposing specific relations among the coefficients of the covariant terms so that the unwanted higher-time-derivative mode is eliminated when the scalar gradient is timelike and the unitary gauge is adopted.  The resulting combinations are therefore obtained through a direct tuning within the covariant ansatz, rather than from an exhaustive classification of the unitary-gauge operator space and its covariant images.  Moreover, the higher-derivative covariant expressions make the degree-of-freedom analysis cumbersome, and their healthy character is tied to the timelike-scalar, unitary-gauge setting.  This motivates a formulation in which temporal and spatial derivatives, and hence potentially dangerous higher-time-derivative structures, are separated from the outset.

Spatially covariant gravity (SCG) provides such a formulation \cite{Gao:2014soa,Gao:2014fra}.  In the unitary gauge of a scalar-tensor theory with a timelike scalar gradient, the scalar defines a preferred foliation and the generally covariant theory is represented by a metric theory invariant under spatial diffeomorphisms.  This unitary-gauge viewpoint is closely related to the effective-field-theory formulations of inflation and dark energy, in which the preferred foliation is used to organize cosmological operators \cite{Cheung:2007st,Gubitosi:2012hu,Gleyzes:2014rba}.  Two advantages of SCG are particularly relevant here.  First, temporal and spatial derivatives are explicit at the operator level, so the lapse velocity, higher normal derivatives of the spatial metric, and the standard first-order-in-time sector can be distinguished before restoring covariance.  Second, SCG provides an efficient constructive route to higher-derivative generally covariant scalar-tensor (GST) theories: one may first organize the operators in unitary gauge and then restore temporal diffeomorphism invariance through the gauge-recovery, or Stueckelberg, procedure.  SCG therefore gives a systematic description of scalar-tensor dynamics rather than merely a collection of Lorentz-violating models.

The correspondence between polynomial SCG and generally covariant GST monomials has been developed through both direct gauge-fixing/gauge-recovery procedures and an explicit linear-algebraic map \cite{Gao:2020juc,Gao:2020yzr,Gao:2020qxy,Hu:2021bbo,Joshi:2021azw,Joshi:2023otx}.  It makes it possible to start from an SCG operator basis with transparent temporal-derivative content and construct its covariant scalar-tensor image while keeping track of the identities induced by the unitary gauge.  The framework has subsequently been generalized by allowing a dynamical lapse, nonmetricity, and auxiliary scalar fields \cite{Gao:2018znj,Gao:2019lpz,Lin:2020nro,Yu:2024drx,Gao:2018izs,Wang:2024hfd}, and it has been applied to cosmological perturbations and gravitational-wave phenomenology \cite{Fujita:2015ymn,Gao:2019liu,Zhu:2022dfq,Zhu:2022uoq,Zhu:2023rrx}. Scalar-induced gravitational waves have also recently been formulated directly within SCG \cite{Jiang:2025ysb}.

Within this programme, polynomial SCG monomials through $d=4$ in the sector without explicit higher normal derivatives (i.e., without the lapse velocity $F$ or the Lie derivative $\mathcal{L}_{\bm{u}}K_{ij}$) have been discussed for both parity-preserving and parity-violating constructions \cite{Gao:2019liu,Gao:2020juc,Gao:2020qxy,Hu:2021bbo}.  Here $d$ denotes the total number of temporal and spatial derivatives in each unitary-gauge monomial.  When the second-order temporal-derivative building blocks $F$ and $\mathcal{L}_{\bm{u}}K_{ij}$ are admitted, the parity-odd SCG monomials through $d=4$ were constructed and classified in Ref.~\cite{Hu:2024hzo}, where their corresponding generally covariant scalar-tensor combinations were also investigated.  The original classification listed one SCG monomial at $d=3$ and eight at $d=4$.  Seven of the nine listed monomials contain neither $F$ nor $\mathcal{L}_{\bm{u}}K_{ij}$ and were mapped to seven generally covariant combinations that avoid the unwanted Ostrogradsky mode in unitary gauge.  As explained in footnote~\ref{fn:z4}, however, one of these seven SCG monomials is algebraically dependent, leaving six independent representatives in this sector.  The remaining two operators already indicate that parity-odd sectors involving higher temporal derivatives require a separate constraint analysis.

In this paper, we generalize the preceding analysis to $d=5$ and perform an exhaustive construction and classification of the polynomial parity-violating SCG monomials at this order.  The extension to total derivative order $d=5$ is motivated by both systematic completeness and phenomenology.  It is the next derivative order at which the parity-odd operator space can be enlarged in a controlled manner, and it provides the necessary SCG input for a future extension of the SCG-GST correspondence to new higher-derivative building blocks.  Moreover, as we show below, monomials at $d=5$ generate new cubic-in-momentum parity-odd contributions to tensor propagation, providing signatures that are absent from the lower-order linear-momentum sector.  A controlled $d=5$ analysis is therefore needed to separate operator classification, degree-of-freedom questions, and the subset of operators directly relevant to gravitational waves.

The construction at $d=5$ is more involved than at lower derivative order.  Candidate products of building blocks and inequivalent index contractions proliferate, third-order building blocks such as $\nabla^2K$, $\nabla R$, and $\nabla\mathcal{L}_{\bm{u}}K$ become necessary, and the sector outside the standard first-order-in-time SCG class expands substantially.  The gravitational-wave analysis introduces a second level of reduction: each independent monomial must be projected onto a homogeneous and isotropic background, expanded to quadratic order in transverse-traceless perturbations, and tested against tensor identities and integrations by parts.  This projection is nontrivial because acceleration operators vanish in the unit-lapse pure-tensor sector, some remaining contractions vanish by the transverse-traceless conditions or reduce to boundary terms, while the genuinely higher-spatial-derivative classes generate new cubic powers of the physical momentum.  Thus the same structures that enlarge the construction at $d=5$ also require a careful separation between the full operator basis and the smaller set of combinations observable in linear tensor propagation.

We accordingly separate the analysis into three steps: the construction of the operator basis at $d=5$, its partition according to temporal-derivative content, and its projection onto the quadratic tensor action.  The reduction yields a $59$-element basis for the parity-odd operator space at $d=5$.  Here ``basis'' means a set of representatives spanning the module of polynomial parity-odd SCG monomials over arbitrary coefficient functions of $(t,N)$, after quotienting by algebraic tensor identities and integrations by parts.  Its temporal-derivative content separates as
\[
59\longrightarrow 35+11+13,
\]
where the three terms denote, respectively, the standard first-order-in-time SCG sector containing neither $F$ nor $\mathcal{L}_{\bm{u}}K_{ij}$, the lapse-velocity sector containing $F$ or its spatial derivatives but no $\mathcal{L}_{\bm{u}}K_{ij}$, and the sector containing at least one $\mathcal{L}_{\bm{u}}K_{ij}$.

We also examine the characteristic imprint of the monomials at $d=5$ on gravitational waves.  For this purpose, the $24$ operators in the lapse-velocity and higher-Lie sectors are set aside because their consistent treatment requires a separate degeneracy and constraint analysis.  On a spatially flat cosmological background, the $19$ acceleration-dependent monomials in the remaining $35$-monomial sector vanish for any homogeneous lapse $N=N(t)$. On the unit-lapse branch used below, this leaves $16$ candidate basis elements in the pure-tensor projection.  Their reduction in the quadratic tensor action can be summarized as
\[
35\longrightarrow 16\longrightarrow 4,
\]
where the last number counts independent time-dependent response combinations rather than individual operators.  Of the $16$ candidates, ten coefficient functions contribute nontrivially and enter only through these four combinations.  We derive the quadratic action in the circular-polarization basis, determine the helicity-dependent kinetic and gradient functions, and identify both the $k/a$ corrections and the $(k/a)^3$ terms specific to $d=5$.  We then discuss the associated amplitude and velocity birefringence, the tensor stability requirements within the effective-field-theory regime, and the conditions under which both helicities propagate luminally.

The paper is organized as follows.  Section~\ref{sec:framework} reviews the SCG building blocks, derivative counting, the lower-order parity-odd bases, and the identities used in the operator reduction.  Section~\ref{sec:classification} constructs and classifies the $d=5$ parity-violating monomials and discusses the temporal-derivative structure and dynamical status of the resulting basis.  Section~\ref{sec:tensor} derives the quadratic tensor action on a cosmological background, explains the projection from the full basis to the tensor-relevant sector, and analyzes birefringence, stability, and luminal propagation.  We conclude in Sec.~\ref{sec:conclusion}, while the complete basis at $d=5$ is collected in Appendix~\ref{app:basis}.

\section{Spatially covariant gravity}
\label{sec:framework}

\subsection{Building blocks of spatially covariant gravity}

Spatially covariant gravity (SCG) is a framework for metric theories invariant under spatial diffeomorphisms \cite{Gao:2014soa,Gao:2014fra}.
The reduction from general covariance to spatial covariance reflects the presence of a preferred foliation of spacetime by spacelike hypersurfaces.
A convenient choice of metric variables is the standard Arnowitt-Deser-Misner variables: the lapse $N$, shift $N^i$, and spatial metric $h_{ij}$, defined through
\begin{equation}
	\mathrm{d} s^2=-N^2\mathrm{d} t^2+h_{ij}(\mathrm{d} x^i+N^i\mathrm{d} t)(\mathrm{d} x^j+N^j\mathrm{d} t).
	\label{eq:admmetric}
\end{equation}
The normal vector to the spacelike hypersurfaces of constant $t$ is denoted by $u^a$, with
\begin{equation}
	(\partial_t)^a=Nu^a+N^a, \label{uaNa}
\end{equation}
where $\partial_{t}$ denotes the time-flow vector field.

In a generally covariant theory, $N$ and $N^i$ implement temporal and spatial diffeomorphisms. In SCG, temporal diffeomorphism invariance is absent, so the lapse $N$ can enter as a physical variable. The basic geometric variables are therefore the normal vector $\bm{u}$ defined by the foliation, the lapse $N$, and the spatial metric $h_{ij}$.
A general SCG action can therefore be written as\footnote{The shift vector enters the Lagrangian only through $\bm{u}$, more precisely through $\mathcal{L}_{\bm{u}}$. This follows because $N^i$ is the gauge field in SCG theories, which appears only when $\bm{u}$ is written explicitly in the adapted coordinates through \eqref{uaNa}.}
\begin{equation}
	S= \int \mathrm{d}t \mathrm{d}^3 x N \sqrt{h} \mathcal{L}(t,N,h_{ij}, R_{ij}, \nabla_{i}, \mathcal{L}_{\bm{u}}, \varepsilon_{ijk}).
\end{equation}
Here $\nabla_i$ is the spatial covariant derivative compatible with $h_{ij}$, and $\mathcal{L}_{\bm{u}}$ is the Lie derivative along $\bm{u}$. The spatial volume form is
\begin{equation}
	\varepsilon_{ijk}=\sqrt{h}\,\epsilon_{ijk},\qquad
	\varepsilon^{ijk}=\frac{\epsilon^{ijk}}{\sqrt{h}}
	\label{eq:epsilonconv}
\end{equation}
with $\epsilon_{123}=\epsilon^{123}=1$.
The intrinsic curvature is described by the three-dimensional Ricci tensor $R_{ij}$ and scalar $R$.

The first-derivative variables are the extrinsic curvature, the velocity of the lapse, and the acceleration,
\begin{align}
	K_{ij}&=\frac{1}{2}\mathcal{L}_{\bm{u}}h_{ij}
	=\frac{1}{2N}(\dot h_{ij}-\nabla_iN_j-\nabla_jN_i),
	\label{eq:Kdef}\\
	F&=\mathcal{L}_{\bm{u}}\ln N
	=\frac{1}{N^2}(\dot N-N^i\nabla_iN),
	\label{eq:Fdef}\\
	a_i&=\nabla_i\ln N.
	\label{eq:adef}
\end{align}
We use the contracted Bianchi identity
\begin{equation}
	\nabla^jR_{ij}=\frac12\nabla_iR,
\end{equation}
and
\begin{equation}
	\nabla_ia_j=\nabla_ja_i,
	\qquad
	\mathcal{L}_{\bm{u}}a_i=\nabla_iF+Fa_i.
	\label{eq:basicidentities}
\end{equation}
Throughout this paper, the Lie derivative $\mathcal{L}_{\bm{u}}$ is defined only for spatial tensors with ``lower'' indices. We choose the convention that Lie derivatives of spatial tensors with ``upper'' indices are merely shorthands of Lie derivatives of spatial tensors with ``lower'' indices after raising the indices. For example, $\mathcal{L}_{\bm{u}}T^{ij}$ denotes $h^{ik}h^{jl}\mathcal{L}_{\bm{u}}T_{kl}$ (instead of first raising the indices and then taking a Lie derivative).

\subsection{Classification of parity-odd SCG monomials}

With this convention, the general action of a SCG theory can be written as
\begin{equation}
	S=\int \mathrm{d} t\,\mathrm{d}^3x\,N\sqrt{h}\,
	\mathcal{L}(t,N,h_{ij},K_{ij},F,a_i,R_{ij},\nabla_i,\mathcal{L}_{\bm{u}},\varepsilon_{ijk}).
	\label{eq:generalaction}
\end{equation}
In this work, we focus on SCG theories with polynomial Lagrangians. 
Specifically, the Lagrangian is a combination of monomials built from these building blocks. The coefficients are assumed to be functions of $t$ and $N$.  Their derivatives satisfy
\begin{align}
	\nabla_if(t,N)&=Nf_{,N}a_i, \label{eq:coeffnabla}\\
	\mathcal{L}_{\bm{u}}f(t,N)&=\frac{1}{N}f_{,t}+Nf_{,N}F.
	\label{eq:coeffderivs}
\end{align}

We classify both the building blocks and the monomials constructed from them according to the numbers of temporal and spatial derivatives they contain.
The basic variables $N$ and $h_{ij}$ have no derivative.
We assign one temporal derivative to each factor of $K_{ij}$ or $F$, one spatial derivative to $a_i$, and two spatial derivatives to $R_{ij}$ and $\nabla_ia_j$. Each additional action of $\mathcal{L}_{\bm{u}}$ increases the temporal-derivative count by one, whereas each additional spatial covariant derivative $\nabla_i$ increases the spatial-derivative count by one. We denote the numbers of temporal and spatial derivatives in each monomial as $d_{\mathrm{t}}$ and $d_{\mathrm{s}}$, respectively. 
For example,
\begin{equation}
	\varepsilon^{ijk}K_i{}^lR_{jm}\nabla_kK_l{}^m
	\quad\hbox{has}\quad (d_{\mathrm{t}},d_{\mathrm{s}})=(2,3),
\end{equation}
and
\begin{equation}
	\varepsilon^{ijk}K^{lm}K_{li}K_j{}^n\nabla_kK_{mn}
	\quad\hbox{has}\quad (d_{\mathrm{t}},d_{\mathrm{s}})=(4,1).
\end{equation}

The total derivative order of a monomial is therefore
\begin{equation}
	d=d_{\mathrm{t}}+d_{\mathrm{s}}.
	\label{eq:ddef}
\end{equation}
In \cite{Gao:2020juc}, the integers $(c_0,\cdots;d_{2},d_{3},\cdots)$ provide a finer classification of generally covariant scalar-tensor monomials and their SCG counterparts. 
Because a parity-odd SCG monomial contains a Levi-Civita tensor $\varepsilon_{ijk}$, $d_{\mathrm{s}}$ must be odd.
The number $d$ in \eqref{eq:ddef} coincides with the integer $d$ in \cite{Gao:2020juc} defined by
\[
d = \sum_{n=0} [(n+2)c_n+(n+1)d_{n+2}].
\]
See Refs.~\cite{Gao:2020juc,Gao:2020yzr} for further details.
In an effective-field-theory interpretation, $d$ is also the derivative, or mass, dimension of the monomial.

Monomials can also be related by integrations by parts. Spatial integration by parts preserves $(d_{\text{t}},d_{\text{s}})$, modulo the redefinition of the coefficient functions. Schematically,
\[
A\nabla_{i}B\simeq B\nabla_{i}A\oplus a_{i}AB,
\]
where $\simeq$ denotes equality up to a boundary term and $\oplus$ denotes a linear combination with redefined coefficient functions.
Integrations by parts along $\bm{u}$ are more involved because the derivatives of $N\sqrt{h}$ and of the coefficient functions must also be included. For parity-odd monomials, the Levi-Civita tensor and the Schouten identity simplify integrations by parts involving Lie derivatives. Schematically,
\[
\varepsilon^{ijk}A_{i}\mathcal{L}_{\bm{u}}Y_{jk}\simeq\varepsilon^{ijk}\mathcal{L}_{\bm{u}}A_{i}Y_{jk}\oplus F\varepsilon^{ijk}A_{i}Y_{jk}\oplus\varepsilon^{ijk}A_{i}Y_{jk}.
\]
The last term on the right-hand side contains one fewer temporal derivative than the original monomial and therefore belongs to a lower-derivative sector. It does not affect the classification at fixed $(d_{\mathrm{t}},d_{\mathrm{s}})$.

Complete bases for the parity-odd SCG monomials up to $d=4$ have been discussed in \cite{Hu:2024hzo}.
For $d=3$, there is only one monomial with $(d_{\mathrm{t}},d_{\mathrm{s}})=(2,1)$. The complete basis of parity-violating SCG polynomials at $d=3$ is therefore
\begin{equation}
	\mathcal{Z}_{3} = \left\{ \varepsilon_{ijk}K_{l}^{i}\nabla^{j}K^{kl}\right\}. \label{calZ3}
\end{equation}
For $d=4$, there are four independent monomials with $(d_{\mathrm{t}},d_{\mathrm{s}})=(3,1)$ and three with $(d_{\mathrm{t}},d_{\mathrm{s}})=(1,3)$.
The basis for the parity-violating SCG polynomials for $d=4$ is given by\footnote{Ref. \cite{Hu:2024hzo} included three $[KK\nabla K]$ monomials in $\mathcal{Z}_4$, leading to $\dim\mathcal{Z}_4=8$. However, only two of these three monomials are algebraically independent. Defining $M_{1}\equiv\varepsilon_{ijk}K^{im}K^{jn}\nabla_{m}K^{k}_{n}$, $M_{2}\equiv\varepsilon_{ijk}K^{mn}K^{i}_{m}\nabla^{j}K^{k}_{n}$, and $M_{3}\equiv\varepsilon_{ijk}KK^{i}_{l}\nabla^{j}K^{kl}$, one finds the identity $M_{1}-M_{2}+M_{3}=0$. We therefore choose $M_1$ and $M_2$ as the independent representatives in \eqref{calZ4}.\label{fn:z4}}
\begin{eqnarray}
	\mathcal{Z}_{4} & = & \Big\{\varepsilon_{ijk}R_{l}^{i}K^{jl}a^{k},\varepsilon_{ijk}K^{im}K^{jn}\nabla_{m}K_{n}^{k},\varepsilon_{ijk}K^{mn}K_{m}^{i}\nabla^{j}K_{n}^{k},\varepsilon_{ijk}K_{l}^{i}a^{j}\nabla^{k}a^{l},\nonumber \\
	&  & \varepsilon_{ijk}R_{l}^{i}\nabla^{j}K^{kl},\varepsilon_{ijk}a^{i}K^{jl}\mathcal{L}_{\bm{u}}K_{l}^{k},\varepsilon_{ijk}FK_{l}^{i}\nabla^{j}K^{kl}\Big\}. \label{calZ4}
\end{eqnarray}
As a result, the parity-odd basis has dimensions
\begin{equation}
	\dim\mathcal Z_3=1,\qquad \dim\mathcal Z_4=7.
	\label{eq:lowercount}
\end{equation}

As noted above, $d_{\mathrm{s}}$ must be odd for parity-violating SCG monomials.
Hence, at $d=5$, the parity-odd SCG monomials fall into three sectors:
\begin{equation}
	(d_{\mathrm{t}},d_{\mathrm{s}})=(0,5),\qquad (2,3),\qquad (4,1).
	\label{eq:sectors}
\end{equation}
At $d=5$, it is sufficient to retain building blocks of derivative order at most three, because higher-order building blocks can be reduced by integrations by parts.  A convenient inventory is given in Table~\ref{tab:inventory}.  Not every entry survives in the reduced basis. For example, $\mathcal{L}_{\bm{u}}a_i$ can be removed by \eqref{eq:basicidentities}. More generally, the commutator between $\mathcal{L}_{\bm{u}}$ and $\nabla_{i}$ can be employed to reduce the number of independent monomials. Moreover, normal derivatives of intrinsic curvature can be rewritten in terms of spatial derivatives of $K_{ij}$ and $N$. We refer to Appendix \ref{app:math} for details.

The construction at $d=5$ requires not only more contractions but also building blocks carrying two or three derivatives in the counting above.  The displayed basis permits structures quadratic in $\mathcal{L}_{\bm{u}}K_{ij}$, such as $\varepsilon^{ijk}\mathcal{L}_{\bm{u}}K_i{}^l\nabla_j\mathcal{L}_{\bm{u}}K_{kl}$, and contains the purely spatial $R\nabla R$ and higher-spatial-derivative kinetic $\nabla K\nabla^2K$ classes.  The latter two classes generate the cubic-momentum tensor responses derived in Sec.~\ref{sec:tensor}.

\begin{table*}[h]
	\caption{Building blocks relevant to monomials with $d=5$.  The two columns distinguish building blocks associated with the spatial metric from those associated with derivatives of the lapse.}
	\label{tab:inventory}
	\begin{ruledtabular}
		\begin{tabular}{c l l}
			order & metric sector & lapse sector\\
			\hline
			$0$ & $h_{ij}$ & $N$\\
			$1$ & $K_{ij}$ & $a_i,\ F$\\
			$2$ & $\nabla K,\ \mathcal{L}_{\bm{u}}K,\ R$ & $\nabla a,\ \nabla F,\ \mathcal{L}_{\bm{u}}F$\\
			$3$ & $\nabla^2K,\ \nabla\mathcal{L}_{\bm{u}}K,\ \nabla R,\ \mathcal{L}_{\bm{u}}^2K$ & $\nabla^2a,\ \nabla^2F,\ \nabla\mathcal{L}_{\bm{u}}F,\ \mathcal{L}_{\bm{u}}^2F$\\
		\end{tabular}
	\end{ruledtabular}
\end{table*}

Before presenting the explicit classification at $d=5$, we clarify the equivalence relation underlying the construction and summarize the reduction strategy.  At fixed derivative order and derivative partition, candidate monomials are treated as equivalent when they differ by algebraic tensor identities or by integrations by parts, including the induced redefinitions of the arbitrary coefficient functions of $(t,N)$ according to \eqref{eq:coeffnabla} and \eqref{eq:coeffderivs}.  By a basis we mean a set of representatives that is complete and independent with respect to these equivalence relations.  Completeness means that every nonvanishing polynomial parity-odd monomial constructed from the building blocks in Table~\ref{tab:inventory} can be reduced to a linear combination of the retained representatives, while independence means that no nontrivial linear combination of those representatives vanishes identically or reduces to a boundary term under the same relations.

At total derivative order $d=5$, both the number of candidate products of building blocks and the number of inequivalent index contractions grow rapidly, especially in the $(2,3)$ and $(4,1)$ sectors.  We therefore proceed in three stages: first classify the candidates by derivative partition and schematic building-block content, then enumerate the nonvanishing contractions within each schematic class, and finally reduce those contractions to the basis defined above.

Each parity-odd scalar can be represented with one spatial Levi-Civita tensor.  The nonvanishing contractions are reduced using index symmetries, \eqref{eq:basicidentities}, integrations by parts, and the three-dimensional relation between the Riemann and Ricci tensors, 
\begin{equation}
	R^l{}_{mjk}={}R^l{}_jh_{mk}-R^l{}_kh_{mj}
	+\delta^l_jR_{mk}-\delta^l_kR_{mj}
	-\frac{R}{2}(\delta^l_jh_{mk}-\delta^l_kh_{mj}).
	\label{eq:3driemann}
\end{equation}
For contractions involving rank-two tensors, two additional identities are useful.  The Cayley-Hamilton relation for a $(1,1)$ tensor $\bm{M}$ is
\begin{equation}
	\bm{M}^3-I_1\bm{M}^2+I_2\bm{M}-I_3\bm{1}=0,
	\label{eq:CH}
\end{equation}
where $I_1=\mathrm{tr}\bm{M}$, $I_2=[(\mathrm{tr}\bm{M})^2-\mathrm{tr}(\bm{M}^2)]/2$, and $I_3=\det\bm{M}$.  The three-dimensional Schouten identity yields
\begin{equation}
	M^m_{\phantom{m}l}\varepsilon_{ijk}-M^m_{\phantom{m}i}\varepsilon_{ljk}
	+M^m_{\phantom{m}j}\varepsilon_{lik}-M^m_{\phantom{m}k}\varepsilon_{lij}=0.
	\label{eq:schouten}
\end{equation}
Together, these relations provide the algebraic and integration-by-parts reductions used in the next section to identify the retained monomials at $d=5$.


\section{\texorpdfstring{Parity-violating SCG monomials at $d=5$}{Parity-violating SCG monomials at d=5}}
\label{sec:classification}

\subsection{\texorpdfstring{Classification of parity-violating monomials at $d=5$}{Classification of parity-violating monomials at d=5}}

We first organize the monomials at $d=5$ according to how the total number of derivatives is distributed among the constituent building blocks.  Schematically, the relevant partitions can be written as
\begin{align*}
	[d]\equiv [5] & =[1+1+1+1+1]\\
	& =[1+1+1+2]\\
	& =[1+1+3]\\
	& =[1+2+2]\\
	& =[2+3]\\
	& = [1+4].
\end{align*}
Here $[\cdots]$ denotes a schematic product, and each integer is the derivative order carried by one building block according to the counting introduced in Sec.~\ref{sec:framework}.  For example, the partition $[2+3]$ includes the class $[R\nabla R]$, represented by the contraction $\varepsilon^{ijk}R_i{}^l\nabla_jR_{kl}$, since $R_{ij}$ and $\nabla_kR_{ij}$ carry derivative orders two and three, respectively.  Owing to the antisymmetry of the Levi-Civita tensor, the type $[1+1+1+1+1]$ admits no nonzero parity-odd scalar contraction.  The partitions $[1+1+3]$ and $[1+4]$, on the other hand, can be reduced by integrations by parts according to
\begin{align*}
	[1+1+3] & \simeq [1+1+1+2]\oplus [1+2+2],\\
	[1+4] & \simeq [1+1+3]\oplus[2+3]
	\simeq [1+1+1+2]\oplus [1+2+2] \oplus [2+3].
\end{align*}
It therefore suffices to analyze the three irreducible partition types $[1+1+1+2]$, $[1+2+2]$, and $[2+3]$.
At this stage, the partition does not distinguish temporal from spatial derivatives.

Before presenting the explicit construction in each derivative partition, we summarize the counting and notation.  Table~\ref{tab:counting} gives the number of retained monomials, while their explicit expressions are collected in Appendix~\ref{app:basis}.  The letters $U$, $V$, and $W$ label the partitions $(d_{\mathrm{t}},d_{\mathrm{s}})=(0,5)$, $(2,3)$, and $(4,1)$, respectively.  Within each partition, the parenthetical superscripts label families of monomials, which are introduced merely for bookkeeping. The subscript $A$ in, for example, $U^{(1)}_A$ or $V^{(n)}_A$ enumerates the monomials in that family. The dimension of a family is therefore the number of allowed values of $A$.  Some families correspond to a single schematic class with several inequivalent contractions, whereas $U^{(1)}$, $V^{(1)}$, and $W^{(1)}$ collect distinct schematic classes that each contribute one retained contraction.

\begin{table}[h]
	\caption{Parity-violating monomials at total derivative order $d=5$ retained after the reductions described below.  The family labels refer to Appendix~\ref{app:basis}.}
	\label{tab:counting}
	\begin{ruledtabular}
		\begin{tabular}{c c c c}
			$(d_{\mathrm{t}},d_{\mathrm{s}})$ & families & family dimensions & total \\
			\hline
			$(0,5)$ & $U^{(1)}$ & $2$ & $2$ \\
			$(2,3)$ & $V^{(1)},\ldots,V^{(9)}$ & $9,5,2,2,4,6,5,2,5$ & $40$ \\
			$(4,1)$ & $W^{(1)},\ldots,W^{(4)}$ & $5,2,4,6$ & $17$ \\
			\hline
			all & & & $59$
		\end{tabular}
	\end{ruledtabular}
\end{table}

\subsubsection{Purely spatial sector: $(0,5)$}

Following the notation of Ref.~\cite{Hu:2024hzo}, we use $[\cdots]$ to denote a schematic class specified by its building-block content, with all admissible scalar contractions of the tensor indices understood.  The schematic classes admitting at least one nonvanishing contraction are
\begin{equation}
	[a\nabla aR],\quad [R\nabla R],\quad [\nabla a\nabla R],
	\quad [R\nabla^2a],\quad [\nabla a\nabla^2a].
\end{equation}
Here ``nonvanishing'' means that at least one scalar contraction with a single $\varepsilon^{ijk}$ is not identically zero under the immediate index symmetries. 
For example, $[a\nabla aR]$ denotes all inequivalent nonzero scalar contractions constructed from one $a_i$, one $\nabla_j a_k$, one $R_{lm}$, and a single Levi-Civita tensor, with the remaining spatial indices contracted by $h^{ij}$.  At this stage, relations generated by integrations by parts, curvature identities, and derivative commutators have not yet been imposed.
With the common coefficient $f(t,N)$ and the Levi-Civita tensor displayed explicitly, the five classes are represented by
\begin{align*}
	[a\nabla aR]&:\quad f\varepsilon^{ijk}a_i\nabla_ja^lR_{kl},\\
	[R\nabla R]&:\quad f\varepsilon^{ijk}R_i{}^l\nabla_jR_{kl},\\
	[\nabla a\nabla R]&:\quad f\varepsilon^{ijk}\nabla_ia^l\nabla_jR_{kl},\\
	[R\nabla^2a]&:\quad f\varepsilon^{ijk}R_{il}\nabla_j\nabla_ka^l,\\
	[\nabla a\nabla^2a]&:\quad f\varepsilon^{ijk}\nabla_ia_l\nabla_j\nabla_ka^l.
\end{align*}
Before imposing integrations by parts and curvature identities, each of the five schematic classes admits one inequivalent nonzero contraction under the elementary tensor symmetries. 

These five classes can be reduced by integrations by parts and algebraic identities. For example,
\begin{equation}
	Nf\varepsilon^{ijk}\nabla_{i}a^{l}\nabla_{j}R_{kl}\simeq-N\frac{\partial\left(Nf\right)}{\partial N}\varepsilon^{ijk}a_{j}\nabla_{i}a^{l}R_{kl}-Nf\varepsilon^{ijk}\nabla_{j}\nabla_{i}a^{l}R_{kl},
	\label{eq:exampleibp}
\end{equation}
which reduces $[\nabla a\nabla R]$ to the $[a\nabla aR]$ and $[R\nabla^2a]$ classes.  Furthermore, antisymmetrizing the last two derivatives gives
\begin{equation}
	f\varepsilon^{ijk}R_{il}\nabla_j\nabla_ka^l
	=\frac{1}{2}f\varepsilon^{ijk}R_{il}R^l{}_{mjk}a^m,
	\label{eq:RRreduce}
\end{equation}
which vanishes after using \eqref{eq:3driemann} and the tensor symmetries.  Thus $Nf\varepsilon^{ijk}\nabla_{i}a^{l}\nabla_{j}R_{kl}\simeq-N\frac{\partial\left(Nf\right)}{\partial N}\varepsilon^{ijk}a_{j}\nabla_{i}a^{l}R_{kl}$.
At the schematic level, we therefore find
\begin{align*}
	[\nabla a\nabla R]&\simeq [a\nabla aR]\oplus[R\nabla^2a] \rightarrow [a\nabla aR]\oplus0  \rightarrow [a\nabla aR],\\
	[R\nabla^2a]&\rightarrow[aRR]\rightarrow0,\\
	[\nabla a\nabla^2a]&\rightarrow [a\nabla aR].
\end{align*}
The last line follows by antisymmetrizing the two derivatives contracted with $\varepsilon^{ijk}$, using the curvature commutator, and integrating by parts. Only the schematic class relation, rather than its coefficient-level normalization, is required for the counting.

After these reductions, two independent nonvanishing classes remain in the $(0,5)$ sector: $[a\nabla a R]$ and $[R\nabla R]$. Consequently, the two surviving monomials are
\begin{align}
	U^{(1)}_1&=\varepsilon^{ijk}a_i\nabla_ja^lR_{kl},
	\label{eq:U1main}\\
	U^{(1)}_2&=\varepsilon^{ijk}R_i{}^l\nabla_jR_{kl}.
	\label{eq:U2main}
\end{align}
The family $U^{(1)}$ therefore has dimension two, with $A=1,2$ labeling the two retained monomials.

\subsubsection{Mixed sector: $(2,3)$}

The $(2,3)$ sector is the largest.  The schematic classes admitting nonvanishing contractions are
\begin{gather}
	[Kaa\nabla F],\ [KaF\nabla a],\ [Kaa\nabla K],\ [KKa\nabla a],\ [KKaR],\nonumber\\
	[aR\mathcal{L}K],\ [a\nabla a\mathcal{L}K],\ [KaFR],\ [K\nabla FR],\ [FR\nabla K],\ [a\nabla K\nabla K],\nonumber\\
	[KR\nabla K],\ [\mathcal{L}K\nabla R],\ [F\nabla a\nabla K],\ [K\nabla a\nabla K],\ [a\nabla F\nabla K],\ [\nabla K\nabla^2K],\nonumber\\
	[aaF\nabla K],\ [K\nabla F\nabla a],\ [\mathcal{L}K\nabla^2a],\ [\nabla K\nabla^2F],\ [\nabla F\nabla^2K],\ [R\nabla\mathcal{L}K],\ [\nabla a\nabla\mathcal{L}K]. \label{ne23}
\end{gather}
Within these schematic brackets, $\mathcal{L}K$ abbreviates $\mathcal{L}_{\bm{u}}K_{ij}$ with the indices contracted as required.  The seven classes in the last line of \eqref{ne23} are eliminated in favor of the preceding classes.  Modulo boundary terms and coefficient redefinitions, we find the following reduction relations at the class level:
\begin{align*}
	[aaF\nabla K]&\simeq[Kaa\nabla F]\oplus[KaF\nabla a],\\
	[K\nabla F\nabla a]&\simeq[KaF\nabla a]\oplus[F\nabla a\nabla K]\oplus[KaFR],\\
	[\mathcal{L}K\nabla^2a]&\rightarrow[aR\mathcal{L}K],\\
	[\nabla K\nabla^2F]&\simeq[a\nabla F\nabla K]\oplus[K\nabla FR],\\
	[\nabla F\nabla^2K]&\simeq[\nabla K\nabla^2F]\oplus[a\nabla F\nabla K]
	\rightarrow[a\nabla F\nabla K]\oplus[K\nabla FR],\\
	[R\nabla\mathcal{L}K]&\simeq[\mathcal{L}K\nabla R]\oplus[aR\mathcal{L}K],\\
	[\nabla a\nabla\mathcal{L}K]&\simeq[\mathcal{L}K\nabla^2a]\oplus[a\nabla a\mathcal{L}K]
	\rightarrow[aR\mathcal{L}K]\oplus[a\nabla a\mathcal{L}K].
\end{align*}
These relations reduce the schematic classes, after which the contractions within each retained class are further reduced by tensor symmetries and the three-dimensional Cayley-Hamilton and Schouten identities.  The latter identities are particularly important for contractions containing several copies of $K_i{}^j$, $R_i{}^j$, or vectors such as $a_i$.

The reduction leaves nine families in the $(2,3)$ sector.  The first is the single-contraction family $V^{(1)}$: it collects nine retained schematic classes that admit exactly one inequivalent nonzero contraction each.  The resulting monomials are denoted by $V^{(1)}_A$ with $A=1,\ldots,9$, and their explicit expressions are given in \eqref{eq:V11}-\eqref{eq:V19}. Grouping them into one family is only a bookkeeping convention and does not imply any additional equivalence among the nine monomials.

The remaining eight families correspond, respectively, to the schematic classes
\begin{equation}
	[Kaa\nabla K],\quad [KKa\nabla a],\quad [KKaR],\quad
	[a\nabla K\nabla K],\quad [KR\nabla K],\quad
	[K\nabla a\nabla K],\quad [a\nabla F\nabla K],\quad
	[\nabla K\nabla^2K].
\end{equation}
Each of these classes admits more than one inequivalent contraction and defines one of the families $V^{(2)},\cdots,V^{(9)}$ in the order displayed above.  More explicitly, $V^{(2)},\ldots,V^{(9)}$ contain, respectively, $5$, $2$, $2$, $4$, $6$, $5$, $2$, and $5$ inequivalent contractions of the corresponding schematic forms.  The dimensions of all nine families are summarized in Table~\ref{tab:counting}.  The eight multi-contraction families therefore contain
\begin{equation*}
	5+2+2+4+6+5+2+5=31
\end{equation*}
monomials, which together with the nine monomials in $V^{(1)}$ give a total of $40$ monomials in the $(2,3)$ sector.  The explicit expressions of the $31$ monomials in $V^{(2)}$-$V^{(9)}$ are given in \eqref{eq:V21}-\eqref{eq:V95}.

\subsubsection{Predominantly temporal sector: $(4,1)$}

The candidate classes in the $(4,1)$ sector are
\begin{gather}
	[KFF\nabla K],\ [KKF\nabla K],\ [KKK\nabla K],\
	[K\nabla F\mathcal{L} K],\ [F\mathcal{L} K\nabla K],
	\nonumber\\
	[K\mathcal{L} K\nabla K],\ [\mathcal{L} K\nabla\mathcal{L} K],\ [KaF\mathcal{L} K],
	\nonumber\\
	[K\mathcal{L} F\nabla K],\ [KKa\mathcal{L} K],\ [\nabla K\mathcal{L}^2K].
\end{gather}
The last three classes reduce to the preceding classes by temporal and spatial integrations by parts together with the commutators between $\mathcal{L}_{\bm{u}}$ and $\nabla_i$ (see Appendix \ref{app:math}).  

The surviving 17 monomials are organized into four families according to their retained schematic content.  The family $W^{(1)}$ collects five distinct schematic classes that each admit one retained contraction,
\begin{equation}
	[KFF\nabla K],\quad [K\nabla F\mathcal{L}K],\quad
	[F\mathcal{L}K\nabla K],\quad
	[\mathcal{L}K\nabla\mathcal{L}K],\quad
	[KaF\mathcal{L}K].
\end{equation}
The family $W^{(2)}$ contains the two contractions of $[KKF\nabla K]$, $W^{(3)}$ contains the four contractions of $[KKK\nabla K]$, and $W^{(4)}$ contains the six contractions of $[K\mathcal{L}K\nabla K]$.  Their dimensions are therefore $5+2+4+6=17$, as shown in Table~\ref{tab:counting}.  The explicit expressions of all 17 monomials are given in \eqref{eq:W11}-\eqref{eq:W46}.

\subsection{On the temporal derivatives in the basis}
\label{subsec:temporalcontent}

The basis also separates naturally according to its temporal-derivative content.  A direct inspection gives
\begin{equation}
	U^{(1)},\quad V^{(2)},\ldots,V^{(7)},\quad V^{(9)},
	\quad W^{(3)},
	\label{eq:35subset}
\end{equation}
These families comprise $35$ monomials containing neither $F$ nor $\mathcal{L}_{\bm{u}}K_{ij}$.  These monomials depend on the spatial metric through $K_{ij}$ and its spatial derivatives but contain no explicit higher temporal derivatives in the unitary-gauge Lagrangian.  

The remaining $24$ monomials belong to $V^{(1)}$, $V^{(8)}$, $W^{(1)}$, $W^{(2)}$, and $W^{(4)}$.
This $24$-monomial sector is further divided according to the origin of the additional temporal derivatives.  Table~\ref{tab:risk} separates monomials containing $F$ or its spatial derivatives but no $\mathcal{L}_{\bm{u}}K_{ij}$ from those containing at least one $\mathcal{L}_{\bm{u}}K_{ij}$.  Explicitly, $11$ monomials belong to the $F$ sector, while $13$ contain at least one $\mathcal{L}_{\bm{u}}K_{ij}$ and form the $\mathcal{L}K$ sector.  Thus the basis exhibits a $35+11+13$ partition. 

\begin{table*}[h]
	\caption{Classification of the $d=5$ basis according to temporal derivatives.  ``Manifestly first order in time'' denotes monomials containing neither $F$ nor $\mathcal{L}_{\bm{u}}K_{ij}$.  
		The $F$-sector contains $F$ or its spatial derivatives but no $\mathcal{L}_{\bm{u}}K_{ij}$, while the $\mathcal{L}K$-sector contains at least one $\mathcal{L}_{\bm{u}}K_{ij}$.}
	\label{tab:risk}
	\begin{ruledtabular}
		\begin{tabular}{c c c c}
			sector & manifestly first order in time & $F$-sector & $\mathcal{L}K$-sector\\
			\hline
			$(0,5)$ & $2$ & $0$ & $0$\\
			$(2,3)$ & $29$ & $8$ & $3$\\
			$(4,1)$ & $4$ & $3$ & $10$\\
			\hline
			Total & $35$ & $11$ & $13$
		\end{tabular}
	\end{ruledtabular}
\end{table*}

The $35$ monomials in \eqref{eq:35subset} belong to the standard first-order-in-time SCG class: the lapse carries no velocity, and the spatial metric enters with at most one temporal derivative.  Higher spatial derivatives alone do not generate an Ostrogradsky mode associated with higher temporal derivatives \cite{Gao:2014soa,Gao:2014fra}. 

The other two sectors require separate dynamical analyses.  The $11$ monomials in the $F$-sector contain the lapse velocity and may render the lapse dynamical. Their viability therefore depends on whether the full kinetic structure is sufficiently degenerate to remove the additional mode.  The $13$ monomials in the $\mathcal{L}K$-sector contain second-order temporal derivatives of the spatial metric and require a kinetic-matrix and constraint analysis involving $\mathcal{L}_{\bm{u}}K_{ij}$ together with $F$ whenever it is present.  At $d\leq4$, the only parity-odd monomials with these higher-time-derivative structures are the last two terms in \eqref{calZ4}, namely $\varepsilon_{ijk}a^{i}K^{jl}\mathcal{L}_{\bm{u}}K_{l}^{k}$ and $\varepsilon_{ijk}FK_{l}^{i}\nabla^{j}K^{kl}$.  They were identified as potential sources of an Ostrogradsky mode, but the lower-order classification does not establish a healthy nontrivial degenerate combination within this restricted two-operator sector.  At $d=5$, the corresponding sector expands to $24$ monomials, substantially enlarging the space in which cancellations or degeneracy conditions might occur.  Determining whether special combinations are degenerate and propagate the desired number of degrees of freedom requires a full kinetic-matrix and Hamiltonian constraint analysis, which lies beyond the present classification.

\section{Propagation of the gravitational waves}
\label{sec:tensor}

\subsection{Quadratic action for the tensor perturbations}

Tensor perturbations provide a particularly direct probe of parity violation because the two circular polarizations can acquire different kinetic normalizations and propagation speeds.  They also isolate the momentum dependence generated by monomials containing higher spatial derivatives.  In this section, we therefore use the tensor sector to identify the gravitational-wave signatures of the parity-violating SCG monomials at total derivative order $d=5$.

We consider a spatially flat Friedmann-Lema\^{\i}tre-Robertson-Walker (FLRW) background and restrict to transverse-traceless tensor perturbations on the unit-lapse branch,\footnote{Unlike generally covariant theories, SCG has no time-reparametrization gauge symmetry. Its coefficients may depend on $t$ and $N$, and homogeneous profiles $N(t)$ are generally inequivalent.  Thus $N=1$ is a restriction to a unit-lapse branch of solutions to the background equations of motion, not a gauge choice. We use this branch so that $t$ is proper time along the preferred foliation and assume that it satisfies the background equations.}
\begin{equation}
	N=1,\qquad N^i=0,\qquad
	h_{ij}=a^2(t)(e^\gamma)_{ij},
	\qquad \partial_i\gamma^{ij}=0,\quad \gamma^i{}_i=0.
	\label{eq:tensoransatz}
\end{equation}
At quadratic order, scalar, vector, and tensor perturbations decouple by spatial rotational symmetry.  The general quadratic tensor action in SCG around an FLRW background can be written as \cite{Gao:2019liu}
\begin{equation}
	S_2=\frac12\int\mathrm{d} t\,\mathrm{d}^3x\,a^3
	\left(\dot\gamma_{ij}\widehat{\mathcal{G}}^{ij,kl}\dot\gamma_{kl}
	-\gamma_{ij}\widehat{\mathcal{W}}^{ij,kl}\gamma_{kl}\right),
	\label{eq:generalS2}
\end{equation}
where $\widehat{\mathcal{G}}^{ij,kl}$ and $\widehat{\mathcal{W}}^{ij,kl}$ are time-dependent spatial differential operators.  Rotational invariance fixes their general form:
\begin{align}
	\widehat{\mathcal{G}}^{ij,kl}={}&\sum_{n=0}^{\infty}
	\left[\mathcal{G}_{2n}S^{ij,kl}-\frac{1}{a}\mathcal{G}_{2n+1}
	A^{ij,kl,m}\partial_m\right]
	\frac{(-\partial^2)^n}{a^{2n}},
	\label{eq:Ghat}\\
	\widehat{\mathcal{W}}^{ij,kl}={}&\sum_{n=0}^{\infty}
	\left[\mathcal{W}_{2n}S^{ij,kl}-\frac{1}{a}\mathcal{W}_{2n+1}
	A^{ij,kl,m}\partial_m\right]
	\frac{(-\partial^2)^{n+1}}{a^{2n+2}}.
	\label{eq:What}
\end{align}
Here, the coefficients $\mathcal{G}_{n}$ and $\mathcal{W}_{n}$ are functions of $t$, and
\begin{align}
	S^{ij,kl}&=\frac12(\delta^{ik}\delta^{jl}+\delta^{il}\delta^{jk}),\\
	A^{ij,kl,m}&=\frac14(\delta^{ik}\epsilon^{jlm}+\delta^{il}\epsilon^{jkm}
	+\delta^{jk}\epsilon^{ilm}+\delta^{jl}\epsilon^{ikm}).
	\label{eq:SAdefs}
\end{align}

Using conformal time, $\mathrm{d}t=a\,\mathrm{d}\tau$, we decompose the tensor perturbation in momentum space as
\begin{equation}
	\gamma_{ij}(\tau,\bm{k})=
	\sum_{s=\pm2}\gamma^{(s)}(\tau,\bm{k})e^{(s)}_{ij}(\hat{\bm{k}}),
	\label{eq:helicitydecomp}
\end{equation}
with
\begin{equation}
	e^{(s)*}_{ij}(\hat{\bm{k}})=e^{(-s)}_{ij}(\hat{\bm{k}})
	=e^{(s)}_{ij}(-\hat{\bm{k}}),
	\qquad
	e^{(s)}_{ij}e^{(-s')ij}=\delta^{ss'}.
\end{equation}
The polarization tensors $e^{(s)}_{ij}(\hat{\bm{k}})$ satisfy
\begin{equation}
	i\hat k^l\epsilon_{lij}
	e^{(s)i}{}_m(\hat{\bm{k}})e^{(s')jm}(-\hat{\bm{k}})
	=\frac{s}{2}\delta^{ss'}.
	\label{eq:helicityidentity}
\end{equation}

Substituting this decomposition into \eqref{eq:generalS2}, we obtain the quadratic action for the polarization modes,
\begin{align}
	S_2={}&\frac12\int\mathrm{d}\tau\frac{\mathrm{d}^3k}{(2\pi)^3}a^2
	\sum_{s=\pm2}\mathcal{G}^{(s)}(\tau,k)
	\Bigg[\gamma^{(s)\prime}(\bm{k})\gamma^{(s)\prime}(-\bm{k})
	\nonumber\\
	&\hspace{38mm}-k^2\frac{\mathcal{W}^{(s)}(\tau,k)}{\mathcal{G}^{(s)}(\tau,k)}
	\gamma^{(s)}(\bm{k})\gamma^{(s)}(-\bm{k})\Bigg],
	\label{eq:helicityaction}
\end{align}
where
\begin{align}
	\mathcal{G}^{(s)}(\tau,k)&=\sum_{n=0}^{\infty}\mathcal{G}_n(\tau)
	\left(\frac{sk}{2a}\right)^n,
	\label{eq:Gsdef}\\
	\mathcal{W}^{(s)}(\tau,k)&=\sum_{n=0}^{\infty}\mathcal{W}_n(\tau)
	\left(\frac{sk}{2a}\right)^n.
	\label{eq:Wsdef}
\end{align}
Under a parity transformation the two helicities are interchanged, $s\to-s$, whereas $k=|\bm{k}|$ is unchanged.  The terms with even $n$ in \eqref{eq:Gsdef} and \eqref{eq:Wsdef} are therefore invariant and contribute equally to the two helicities, corresponding to parity-preserving propagation.  The terms with odd $n$ reverse sign between the two helicities and encode parity violation. This power-series description parallels general effective parametrizations in which dispersion and birefringence are organized by the momentum dependence of gravitational-wave propagation \cite{Mewes:2019dhj}.

For perturbative stability in the tensor sector one must additionally require
\begin{equation}
	\mathcal{G}^{(s)}(\tau,k)>0,
	\qquad
	\mathcal{W}^{(s)}(\tau,k)>0
	\label{eq:tensorstability}
\end{equation}
throughout the momentum range in which the derivative expansion is valid. 

\subsection{\texorpdfstring{Contributions to gravitational-wave propagation}{Contributions to gravitational-wave propagation}}

The basis at $d=5$ contains 59 monomials.  To isolate tensor signatures without first resolving the degeneracy problem of the full theory, we restrict the present calculation to the sector that is manifestly first order in temporal derivatives.  Monomials containing $\mathcal L_{\bm u}K_{ij}$ are set aside because this building block contains second time derivatives of $h_{ij}$, and hence of the tensor perturbations $\gamma_{ij}$. Their consistent perturbative treatment requires a separate kinetic-matrix and constraint analysis.  Monomials containing $F$ also do not contribute to the strict unit-lapse tensor projection used here.\footnote{For the ansatz \eqref{eq:tensoransatz}, $N=1$, $N^i=0$, and no lapse perturbation is retained.  Consequently, $F=0$ and all of its spatial derivatives vanish, so monomials containing either $F$ or $\nabla_iF$ do not contribute.  More generally, on a homogeneous branch with $N=N(t)$, spatial derivatives of $F$ still vanish, whereas the background value $\bar F=\dot{\bar N}/\bar N^2$ need not.  Monomials containing an undifferentiated $F$, such as $W^{(1)}_{1}$ in \eqref{eq:W11}, may therefore contribute to tensor perturbations on such a branch. \label{fn:F}}  We therefore set aside all 24 monomials containing $F$ and/or $\mathcal L_{\bm u}K_{ij}$.  Of the remaining 35 monomials in \eqref{eq:35subset}, 19 contain at least one factor of $a_i$ or its spatial derivatives.  Since $a_i=\nabla_i\ln N=0$ on any homogeneous FLRW background with $N=N(t)$, these monomials do not contribute to linear tensor perturbations.  The resulting 16 monomials are listed in \eqref{eq:LPVtensor} and fall into the four families summarized in Table~\ref{tab:tensorresponses}.  This selection isolates a controlled tensor sector at $d=5$. It does not imply that the omitted higher-time-derivative monomials are dynamically irrelevant.

At quadratic order, the 16 selected monomials enter through only four response combinations.  The terms proportional to the physical momentum $k/a$ in $\mathcal G^{(s)}$ and $\mathcal W^{(s)}$ have the same momentum scaling as the parity-odd tensor corrections already possible at $d\leq4$, although here they are generated by the $d=5$ classes $[KKK\nabla K]$ and $[KR\nabla K]$.  By contrast, the terms proportional to $(k/a)^3$ are absent in the $d\leq4$ truncation and first arise at $d=5$ from the higher-spatial-derivative classes $[\nabla K\nabla^2K]$ and $[R\nabla R]$.  Since the gradient part of the action carries an overall factor of $k^2$, the cubic term in $\mathcal W^{(s)}$ yields a parity-odd contribution with five powers of spatial momentum, as expected from the purely spatial $d=5$ class $[R\nabla R]$.

\begin{table*}[h]
	\caption{Contributions from the $16$ tensor-relevant basis elements to the four coefficient combinations.  ``Kinetic'' and ``gradient'' refer to $\mathcal G^{(s)}$ and $\mathcal W^{(s)}$, respectively.}
	\label{tab:tensorresponses}
	\begin{ruledtabular}
		\begin{tabular}{l l c c}
			monomial family & coefficient combination & contribution & momentum order\\
			\hline
			$[KKK\nabla K]$ & $\alpha_1=-\bar w^{(3)}_1-\bar w^{(3)}_2+2\bar w^{(3)}_3+3\bar w^{(3)}_4$ & kinetic & $k/a$\\
			$[\nabla K\nabla^2K]$ & $\alpha_3=\bar v^{(9)}_4-\bar v^{(9)}_5$ & kinetic & $(k/a)^3$\\
			$[KR\nabla K]$ & $\beta_1=\bar v^{(6)}_3+\bar v^{(6)}_4-\bar v^{(6)}_6$ & gradient & $k/a$\\
			$[R\nabla R]$ & $\beta_3=-\bar u^{(1)}_2$ & gradient & $(k/a)^3$
		\end{tabular}
	\end{ruledtabular}
\end{table*}

We consider the action
\begin{equation}
	S=\int\mathrm{d}t\,\mathrm{d}^3x\,N\sqrt{h}
	\left(\mathcal{L}_{\rm GR}+\mathcal{L}_{\rm PV}+\mathcal{L}_{\mathrm{m}}\right). \label{model}
\end{equation}
We take general relativity as the parity-even reference theory and set $M_{\mathrm{pl}}^2=1$,
\begin{equation}
	\mathcal{L}_{\rm GR}=\frac12(R-K^2+K_{ij}K^{ij}),
	\label{eq:LGR}
\end{equation}
and restrict the parity-odd sector at $d=5$ to the acceleration-free monomials containing neither $F$ nor $\mathcal{L}_{\bm{u}}K_{ij}$ that survive the pure tensor projection,
\begin{equation}
	\mathcal{L}_{\rm PV}=u^{(1)}_{2}U^{(1)}_2
	+\sum_{A=1}^{6} v^{(6)}_A V^{(6)}_A
	+\sum_{A=1}^{5} v^{(9)}_A V^{(9)}_A
	+\sum_{A=1}^{4} w^{(3)}_A W^{(3)}_A.
	\label{eq:LPVtensor}
\end{equation}
All coefficients in \eqref{eq:LPVtensor} are functions of $t$ and $N$. We retain only parity-odd monomials at $d=5$ in order to isolate their contributions to tensor propagation.  Parity-odd monomials at lower values of $d$ can be added independently. Their classification and effects on tensor propagation have been studied in Refs.~\cite{Gao:2019liu,Hu:2024hzo}.  They would add further terms proportional to $k/a$ to the same response functions, but would not alter the identification of the $(k/a)^3$ terms as effects specific to $d=5$. Note that in \eqref{model} we also include a matter sector $\mathcal{L}_{\mathrm{m}}$ to support the cosmological background, assuming that its tensor anisotropic stress vanishes at linear order so that it affects the tensor sector only through the background evolution. 

The omitted higher-time-derivative sectors behave differently under the tensor projection.  As argued in footnote \ref{fn:F}, the 11 monomials containing $F$ or its spatial derivatives but no $\mathcal L_{\bm u}K_{ij}$ do not contribute to linear tensor perturbations (under the unit-lapse assumption).  If scalar perturbations are taken into account, however, $N=1+\delta N$ and $F$ is generically nonzero at linear order.  These monomials may then enter the scalar constraint system, although determining which of them contributes nontrivially to the quadratic scalar action requires a dedicated perturbative analysis.  Among the 13 monomials containing at least one $\mathcal L_{\bm u}K_{ij}$, those not multiplied by a vanishing $F$ factor may contribute directly to tensor perturbations.  Since $\mathcal L_{\bm u}K_{ij}$ contains a second time derivative of the spatial metric, a nondegenerate quadratic action involving these terms can propagate additional tensor degrees of freedom and is generically expected to exhibit an Ostrogradsky instability.  Degeneracy conditions may remove the extra mode, but establishing this requires the full kinetic matrix and constraint analysis, which is outside the scope of the current work.

A bar denotes evaluation of a coefficient on the unit-lapse background.  Evaluating \eqref{eq:LPVtensor} to quadratic order, we find
\begin{align}
	\mathcal{G}^{(s)}={}&\frac14
	+\frac{\mathcal{H}^2}{2a^2}
	\left(-\bar w^{(3)}_1-\bar w^{(3)}_2
	+2\bar w^{(3)}_3+3\bar w^{(3)}_4\right)
	\left(\frac{sk}{2a}\right)
	\nonumber\\
	&+\frac12(\bar v^{(9)}_4-\bar v^{(9)}_5)
	\left(\frac{sk}{2a}\right)^3,
	\label{eq:Gresult}\\
	\mathcal{W}^{(s)}={}&\frac14
	+\frac{1}{4a}\partial_\tau\left[
	\frac{\mathcal{H}}{a}(\bar v^{(6)}_3+\bar v^{(6)}_4-\bar v^{(6)}_6)
	\right]\left(\frac{sk}{2a}\right)
	\nonumber\\
	&-\frac12\bar u^{(1)}_2
	\left(\frac{sk}{2a}\right)^3.
	\label{eq:Wresult}
\end{align}
Here $\mathcal{H}=a'/a$.  For later use, we define
\begin{align}
	\alpha_1&=-\bar w^{(3)}_1-\bar w^{(3)}_2
	+2\bar w^{(3)}_3+3\bar w^{(3)}_4,
	\label{eq:alpha1}\\
	\alpha_3&=\bar v^{(9)}_4-\bar v^{(9)}_5,
	\label{eq:alpha3}\\
	\beta_1&=\bar v^{(6)}_3+\bar v^{(6)}_4-\bar v^{(6)}_6,
	\label{eq:beta1}\\
	\beta_3&=-\bar u^{(1)}_2.
	\label{eq:beta3}
\end{align}

Within the 16 monomials in \eqref{eq:LPVtensor}, $V^{(6)}_{1,2}$ and $V^{(9)}_{1,2}$ vanish by the transverse-traceless conditions, while $V^{(6)}_5$ and $V^{(9)}_3$ yield no independent quadratic contribution after integration by parts.  Consequently, ten coefficient functions contribute nontrivially to the quadratic tensor action, and they enter only through the four combinations $\alpha_1$, $\alpha_3$, $\beta_1$, and $\beta_3$.

The mode equation is
\begin{equation}
	\gamma^{(s)\prime\prime}
	+\mathcal{H}\left(2+\nu^{(s)}\right)\gamma^{(s)\prime}
	+(c_T^{(s)})^2k^2\gamma^{(s)}=0,
	\label{eq:modeeq}
\end{equation}
with
\begin{equation}
	\nu^{(s)}=\frac{1}{\mathcal{H}}
	\frac{\partial_\tau\mathcal{G}^{(s)}}{\mathcal{G}^{(s)}},
	\qquad
	(c_T^{(s)})^2=\frac{\mathcal{W}^{(s)}}{\mathcal{G}^{(s)}}.
	\label{eq:nuct}
\end{equation}
Defining $x_s\equiv sk/(2a)$, so that $x_s'=-\mathcal Hx_s$ at fixed comoving momentum, the explicit expressions are
\begin{align}
	\nu^{(s)}={}&
	\frac{
		\left[
		\partial_\tau\left(\frac{\mathcal H^2}{2a^2}\alpha_1\right)
		-\mathcal H\frac{\mathcal H^2}{2a^2}\alpha_1
		\right]x_s
		+\frac12\left(\alpha_3'-3\mathcal H\alpha_3\right)x_s^3
	}
	{\mathcal H\left[
		\frac14+\frac{\mathcal H^2}{2a^2}\alpha_1x_s
		+\frac12\alpha_3x_s^3\right]},\\
	(c_T^{(s)})^2={}&
	\frac{
		\frac14+\frac{1}{4a}\partial_\tau\left(\frac{\mathcal H}{a}\beta_1\right)x_s
		+\frac12\beta_3x_s^3
	}
	{
		\frac14+\frac{\mathcal H^2}{2a^2}\alpha_1x_s
		+\frac12\alpha_3x_s^3
	}.
\end{align}
The odd powers of $x_s$ reverse sign between the two helicities.  The combinations $\alpha_1$ and $\alpha_3$ therefore produce a helicity-dependent kinetic normalization and, generically, a helicity-dependent damping rate $\nu^{(s)}$.  A mismatch between the corresponding kinetic and gradient coefficients produces velocity birefringence through the explicit ratio above.

The $d=5$ classes $[KKK\nabla K]$ and $[KR\nabla K]$ add contributions proportional to $k/a$, with the same momentum scaling as lower-order parity-odd terms but a different background and coupling dependence.  The distinctive $d=5$ effect is that $[\nabla K\nabla^2K]$ and $[R\nabla R]$ generate new $(k/a)^3$ contributions governed by $\alpha_3$ and $\beta_3$.  A nonzero $\alpha_3$ therefore produces a helicity-dependent cubic contribution to the kinetic function and can generate amplitude birefringence.  Cubic-momentum velocity birefringence is controlled instead by the mismatch between the kinetic and gradient responses. To first order in the parity-odd corrections, the cubic contribution to $(c_T^{(s)})^2=\mathcal W^{(s)}/\mathcal G^{(s)}$ is proportional to $(\beta_3-\alpha_3)x_s^3$, which reverses sign between the two helicities.  When $\alpha_3=\beta_3$, the direct cubic terms in the numerator and denominator coincide and cancel from their ratio at this order.  This condition alone does not enforce luminal propagation, because the terms linear in $x_s$ may still differ. Moreover, $\mathcal G^{(s)}$ can remain helicity dependent, so amplitude birefringence can persist.
These helicity-dependent amplitude and phase corrections are the quantities targeted by waveform-based and model-independent tests of parity symmetry in compact-binary signals \cite{Qiao:2019wsh,Zhao:2019szi}.

\subsection{Luminal propagation}
\label{subsec:luminal}

Reference~\cite{Gao:2019liu} showed that spatially covariant gravity contains nontrivial parity-violating subclasses for which both tensor helicities propagate at the speed of light.  Such theories can retain parity-dependent kinetic normalization while satisfying the stringent observational bound on the gravitational-wave speed.  We now determine whether the monomials at $d=5$ admit the same property.

From \eqref{eq:nuct}, luminal propagation requires $\mathcal W^{(s)}=\mathcal G^{(s)}$.  The functions in \eqref{eq:Gresult} and \eqref{eq:Wresult} contain the independent odd powers $x_s$ and $x_s^3$.  Requiring equality for both helicities and for every momentum within the $d=5$ truncation therefore requires the coefficients of these two powers to agree separately.  The appearance of two conditions is due to the two independent momentum powers, rather than to the time variation of $\mathcal H$.  The resulting fixed-background conditions are
\begin{align}
	2\mathcal{H}^2\alpha_1&=
	a\partial_\tau\left(\frac{\mathcal{H}}{a}\beta_1\right),
	\label{eq:luminal1}\\
	\bar u^{(1)}_2+\bar v^{(9)}_4-\bar v^{(9)}_5&=0.
	\label{eq:luminal3}
\end{align}
The second relation is equivalently $\alpha_3=\beta_3$.

For a fixed background, the first condition matches the background-dressed $k/a$ kinetic and gradient responses, while the second directly relates the two $(k/a)^3$ classes.  These conditions constrain the phase velocity but do not generally remove the helicity dependence of $\mathcal G^{(s)}$. A parity-dependent damping rate may therefore remain. 

The first relation \eqref{eq:luminal1} can be written more explicitly as
\begin{equation}
	a\partial_\tau\left(\frac{\mathcal H}{a}\beta_1\right)
	=\beta_1\mathcal H'+\mathcal H\beta_1'
	-\mathcal H^2\beta_1,
\end{equation}
and hence
\begin{equation}
	\beta_1\mathcal H'+\mathcal H\beta_1'
	-\mathcal H^2\left(\beta_1+2\alpha_1\right)=0.
\end{equation}
For a specified cosmological background, this is a single differential relation between $\alpha_1$ and $\beta_1$ and does not split into further algebraic constraints.  A stronger result follows only if luminality is required to hold identically for arbitrary expansion histories, with the coupling combinations treated as prescribed functions of $t$ (after setting $N=1$) rather than as functionals of $\mathcal H$ and $\mathcal H'$.  In that case the coefficient of $\mathcal H'$ requires $\beta_1=0$ identically, so that $\beta_1'=0$ as well.  The remaining $\mathcal H^2$ term then requires $\alpha_1=0$.  Together with the cubic-momentum condition, this gives
\begin{equation}
	\alpha_1=0,\qquad \beta_1=0,\qquad \alpha_3=\beta_3.
\end{equation}
In terms of the original coefficients, the background-independent conditions are
\begin{align}
	-\bar w^{(3)}_1-\bar w^{(3)}_2
	+2\bar w^{(3)}_3+3\bar w^{(3)}_4&=0,\\
	\bar v^{(6)}_3+\bar v^{(6)}_4-\bar v^{(6)}_6&=0,\\
	\bar u^{(1)}_2+\bar v^{(9)}_4-\bar v^{(9)}_5&=0.
\end{align}
These stronger conditions are not required when the theory is evaluated only on a chosen cosmological solution.

The luminality conditions constrain only the tensor phase velocity.  They do not eliminate the helicity dependence of $\mathcal{G}^{(s)}$.  Therefore, unless the odd kinetic combinations vanish or evolve in a specially tuned way, \eqref{eq:luminal1}-\eqref{eq:luminal3} allow luminal propagation together with amplitude birefringence.  This is the $d=5$ counterpart of the broader SCG mechanism identified at lower derivative order \cite{Gao:2019liu}.

\section{Conclusion}
\label{sec:conclusion}

We have constructed and organized the polynomial parity-violating SCG monomials at total derivative order $d=5$.  After applying the algebraic identities and integration-by-parts relations used in the present calculation, the resulting set provides a $59$-element basis, consisting of two monomials in the $(d_{\mathrm{t}},d_{\mathrm{s}})=(0,5)$ sector, $40$ in the $(2,3)$ sector, and $17$ in the $(4,1)$ sector.  According to their temporal-derivative content, these monomials further separate into $35$ monomials containing neither the lapse velocity nor a Lie derivative of the extrinsic curvature, $11$ lapse-velocity monomials, and $13$ monomials containing at least one $\mathcal{L}_{\bm{u}}K_{ij}$.  The latter two sectors require a dedicated degeneracy and constraint analysis, which lies beyond the scope of the present work.

We have also investigated the linear propagation of tensor perturbations on a spatially flat cosmological background.  Restricting to the acceleration-free part of the manifestly first-order-in-time sector leaves $16$ basis elements belonging to the $[R\nabla R]$, $[KR\nabla K]$, $[\nabla K\nabla^2K]$, and $[KKK\nabla K]$ families.  At quadratic order in tensor perturbations, their ten coefficient functions enter through only four independent combinations.  These combinations generate helicity-odd corrections proportional to $k/a$ and $(k/a)^3$ in the kinetic and gradient functions.  Requiring both helicities to propagate luminally yields the two relations \eqref{eq:luminal1} and \eqref{eq:luminal3}. Even when these conditions hold, a parity-dependent kinetic normalization, and hence amplitude birefringence, may remain.

The construction at $d=5$ reveals a hierarchy between the complete SCG operator space, the sectors whose consistency depends on new degeneracy conditions, and the much smaller set of combinations that controls linear tensor propagation.  In particular, the $(k/a)^3$ terms provide parity-odd contributions specific to $d=5$ to gravitational-wave propagation, whereas the full tensor response is encoded by only four time-dependent combinations.

A direct next step is to extend the SCG-GST correspondence to total derivative order $d=5$.  As reviewed in the Introduction, generally covariant expressions can be recovered from SCG operators through the Stueckelberg, or gauge-recovery, procedure \cite{Gao:2020yzr,Gao:2020qxy}, while the covariant $3+1$ correspondence of Ref.~\cite{Hu:2021bbo} provides an appropriate framework for examining degeneracy without imposing unitary gauge.  For $d\leq4$, the GST monomial space was constructed in Ref.~\cite{Gao:2020juc}.  The original parity-odd correspondence exhibited seven generally covariant combinations that are degenerate in unitary gauge \cite{Hu:2024hzo}. After accounting for the algebraic relation in footnote~\ref{fn:z4}, six of the corresponding SCG representatives are independent.  The corresponding GST space at total derivative order $d=5$ has not yet been established.  The new SCG structures involving $\nabla R$, $\nabla^2K$, and $\nabla\mathcal{L}_{\bm{u}}K$ are expected to map to covariant expressions containing derivatives of curvature and/or higher covariant derivatives of $\phi$, but this mapping and its degeneracy properties must be derived explicitly.

In the present work, we focused on the $35$ monomials that contain neither $F$ nor $\mathcal{L}_{\bm{u}}K_{ij}$ and are therefore manifestly first order in time in the unitary-gauge Lagrangian. This property excludes an Ostrogradsky instability arising directly from higher temporal derivatives in this sector, but it does not by itself establish complete stability.
The $24$ higher-time-derivative monomials were set aside in the tensor calculation.
A complete dynamical analysis requires construction of the kinetic matrix and a Hamiltonian constraint analysis for the $11+13$ higher-time-derivative sectors.  Such an analysis is necessary for a complete construction of healthy parity-violating scalar-tensor theories at $d=5$.  Moreover, the coefficient dimensions, effective cutoff, background evolution, and initial conditions must be specified before the cubic-momentum terms can be used quantitatively or confronted with gravitational-wave constraints \cite{Gong:2021jgg,Zhang:2025kcw}.
We leave these analyses for future work.

\begin{acknowledgments}
X. G. is supported by the National Natural Science Foundation of China (NSFC) under Grants No. 12475068 and No. 11975020 and the Guangdong Basic and Applied Basic Research Foundation under Grant No. 2025A1515012977.
\end{acknowledgments}

\section*{Data Availability}
No data were created or analyzed in this study.

\appendix

\section{\texorpdfstring{Complete basis for parity-violating SCG monomials at $d=5$}{Complete basis for parity-violating SCG monomials at d=5}}
\label{app:basis}

This appendix collects the explicit representatives of the parity-odd SCG basis at total derivative order $d=5$.  In addition to algebraic tensor identities, the reduction uses temporal and spatial integrations by parts.  These integrations are understood within the polynomial SCG action, where the coefficient multiplying each monomial is an arbitrary function of $t$ and $N$ only.  The basis is organized below according to the derivative split $(d_{\mathrm{t}},d_{\mathrm{s}})$ and the family notation introduced in the main text.

\subsection{The $(0,5)$ sector}

The purely spatial $(0,5)$ sector contains the following two basis elements:
\begin{align}
	U^{(1)}_1&= \varepsilon^{ijk}a_i\nabla_ja^lR_{kl},
	\label{eq:U11}\\
	U^{(1)}_2&= \varepsilon^{ijk}R_i{}^l\nabla_jR_{kl}.
	\label{eq:U12}
\end{align}

\subsection{The $(2,3)$ sector}

The $40$ monomials in the $(2,3)$ sector are organized into nine families.  The first family contains nine monomials arising from schematic classes for which a single inequivalent contraction is retained:
\begin{align}
	V^{(1)}_1&= \varepsilon^{ijk}K_i{}^la_la_j\nabla_kF,
	\label{eq:V11}\\
	V^{(1)}_2&= \varepsilon^{ijk}Fa_iK_j{}^l\nabla_ka_l,\\
	V^{(1)}_3&= \varepsilon^{ijk}a_iR_j{}^l\mathcal{L}_{\bm{u}}K_{kl},\\
	V^{(1)}_4&= \varepsilon^{ijk}a_i\nabla_ja^l\mathcal{L}_{\bm{u}}K_{kl},\\
	V^{(1)}_5&= \varepsilon^{ijk}Fa_iK_j{}^lR_{kl},\\
	V^{(1)}_6&= \varepsilon^{ijk}K_i{}^l\nabla_jF R_{kl},\\
	V^{(1)}_7&= \varepsilon^{ijk}F R_{il}\nabla_jK_k{}^l,\\
	V^{(1)}_8&= \varepsilon^{ijk}\mathcal{L}_{\bm{u}}K_{il}\nabla_jR_k{}^l,\\
	V^{(1)}_9&= \varepsilon^{ijk}F\nabla_ia_l\nabla_jK_k{}^l.
	\label{eq:V19}
\end{align}
The $[Kaa\nabla K]$ family contains five monomials:
\begin{align}
	V^{(2)}_1&= \varepsilon^{ijk}K_i{}^la^ma_j\nabla_kK_{lm}, \label{eq:V21}\\
	V^{(2)}_2&= \varepsilon^{ijk}K_i{}^la_la_j\nabla^mK_{km},\\
	V^{(2)}_3&= \varepsilon^{ijk}K_i{}^la^ma_j\nabla_lK_{km},\\
	V^{(2)}_4&= \varepsilon^{ijk}Ka^la_i\nabla_jK_{kl},\\
	V^{(2)}_5&= \varepsilon^{ijk}a^ma_mK_i{}^l\nabla_jK_{kl}.
	\label{eq:V25}
\end{align}
The $[KKa\nabla a]$ family contains two monomials:
\begin{align}
	V^{(3)}_1&= \varepsilon^{ijk}Ka_iK_j{}^l\nabla_ka_l,\\
	V^{(3)}_2&= \varepsilon^{ijk}K^{lm}K_{il}a_j\nabla_ka_m.
	\label{eq:V32}
\end{align}
The $[KKaR]$ family contains two monomials:
\begin{align}
	V^{(4)}_1&= \varepsilon^{ijk}Ka_iK_j{}^lR_{kl},\\
	V^{(4)}_2&= \varepsilon^{ijk}K^{lm}K_{il}a_jR_{km}.
	\label{eq:V42}
\end{align}
The $[a\nabla K\nabla K]$ family contains four monomials:
\begin{align}
	V^{(5)}_1&= \varepsilon^{ijk}a_i\nabla^lK\nabla_jK_{kl},\\
	V^{(5)}_2&= \varepsilon^{ijk}a_i\nabla_jK\nabla^lK_{kl},\\
	V^{(5)}_3&= \varepsilon^{ijk}a_i\nabla^lK_{lm}\nabla_jK_k{}^m,\\
	V^{(5)}_4&= \varepsilon^{ijk}a_i\nabla_lK_{jm}\nabla_kK^{lm}.
	\label{eq:V54}
\end{align}
The $[KR\nabla K]$ family contains six monomials:
\begin{align}
	V^{(6)}_1&= \varepsilon^{ijk}K_i{}^lR_{jl}\nabla_kK,\\
	V^{(6)}_2&= \varepsilon^{ijk}K_i{}^lR_{jl}\nabla^mK_{km},\\
	V^{(6)}_3&= \varepsilon^{ijk}K_i{}^lR_{lm}\nabla_jK_k{}^m,\\
	V^{(6)}_4&= \varepsilon^{ijk}K_i{}^lR_{jm}\nabla_kK_l{}^m,\\
	V^{(6)}_5&= \varepsilon^{ijk}K_i{}^lR_{jm}\nabla^mK_{kl},\\
	V^{(6)}_6&= \varepsilon^{ijk}K_i{}^lR_{jm}\nabla_lK_k{}^m.
	\label{eq:V66}
\end{align}
The $[K\nabla a\nabla K]$ family contains five monomials:
\begin{align}
	V^{(7)}_1&= \varepsilon^{ijk}K_i{}^l\nabla_la_m\nabla_jK_k{}^m,\\
	V^{(7)}_2&= \varepsilon^{ijk}K_i{}^l\nabla_ja_l\nabla^mK_{km},\\
	V^{(7)}_3&= \varepsilon^{ijk}K_i{}^l\nabla_ja_m\nabla_lK_k{}^m,\\
	V^{(7)}_4&= \varepsilon^{ijk}K_i{}^l\nabla_ja_m\nabla^mK_{kl},\\
	V^{(7)}_5&= \varepsilon^{ijk}K_i{}^l\nabla_ma^m\nabla_jK_{kl}.
	\label{eq:V75}
\end{align}
The $[a\nabla F\nabla K]$ family contains two monomials:
\begin{align}
	V^{(8)}_1&= \varepsilon^{ijk}a_i\nabla_jF\nabla^lK_{kl},\\
	V^{(8)}_2&= \varepsilon^{ijk}a_i\nabla^lF\nabla_jK_{kl}.
	\label{eq:V82}
\end{align}
The ninth family, $[\nabla K\nabla^2K]$, contains five monomials:
\begin{align}
	V^{(9)}_1&= \varepsilon^{ijk}\nabla^lK_{il}\nabla^m\nabla_jK_{km},\\
	V^{(9)}_2&= \varepsilon^{ijk}\nabla_iK_j{}^l\nabla_l\nabla^mK_{km},\\
	V^{(9)}_3&= \varepsilon^{ijk}\nabla^lK_i{}^m\nabla_m\nabla_jK_{kl},\\
	V^{(9)}_4&= \varepsilon^{ijk}\nabla^lK_i{}^m\nabla_l\nabla_jK_{km},\\
	V^{(9)}_5&= \varepsilon^{ijk}\nabla_iK_j{}^l\nabla^m\nabla_mK_{kl}.
	\label{eq:V95}
\end{align}

Together with the nine elements of $V^{(1)}$, the remaining eight families contain $31$ monomials, giving $40$ basis elements in the $(2,3)$ sector.

\subsection{The $(4,1)$ sector}

The $17$ monomials in the $(4,1)$ sector are organized into four families.  Analogously to $V^{(1)}$, the first family contains five monomials arising from schematic classes for which a single inequivalent contraction is retained:
\begin{align}
	W^{(1)}_1&= \varepsilon^{ijk}F^2K_i{}^l\nabla_jK_{kl},
	\label{eq:W11}\\
	W^{(1)}_2&= \varepsilon^{ijk}K_i{}^l\nabla_jF\mathcal{L}_{\bm{u}}K_{kl},\\
	W^{(1)}_3&= \varepsilon^{ijk}F\mathcal{L}_{\bm{u}}K_{il}\nabla_jK_k{}^l,\\
	W^{(1)}_4&= \varepsilon^{ijk}\mathcal{L}_{\bm{u}}K_i{}^l\nabla_j\mathcal{L}_{\bm{u}}K_{kl},\\
	W^{(1)}_5&= \varepsilon^{ijk}Fa_iK_j{}^l\mathcal{L}_{\bm{u}}K_{kl}.
	\label{eq:W15}
\end{align}
The $[KKF\nabla K]$ family contains two monomials:
\begin{align}
	W^{(2)}_1&= \varepsilon^{ijk}F K_i{}^lK_j{}^m\nabla_lK_{km},\\
	W^{(2)}_2&= \varepsilon^{ijk}F K_{il}K^{lm}\nabla_jK_{km}.
	\label{eq:W22}
\end{align}
The $[KKK\nabla K]$ family contains four monomials:
\begin{align}
	W^{(3)}_1&= \varepsilon^{ijk}K^{lm}K_{li}K_j{}^n\nabla_kK_{mn},\\
	W^{(3)}_2&= \varepsilon^{ijk}K^{lm}K_{li}K_j{}^n\nabla_mK_{kn},\\
	W^{(3)}_3&= \varepsilon^{ijk}K^{lm}K_{li}K_j{}^n\nabla_nK_{km},\\
	W^{(3)}_4&= \varepsilon^{ijk}K^{lm}K_{mn}K_i{}^n\nabla_jK_{kl}.
	\label{eq:W34}
\end{align}
The $[K\mathcal{L} K\nabla K]$ family contains six monomials:
\begin{align}
	W^{(4)}_1&= \varepsilon^{ijk}K_i{}^l\mathcal{L}_{\bm{u}}K_{jl}\nabla_kK,\\
	W^{(4)}_2&= \varepsilon^{ijk}K_i{}^l\mathcal{L}_{\bm{u}}K_{jl}\nabla^mK_{km},\\
	W^{(4)}_3&= \varepsilon^{ijk}K_i{}^l\mathcal{L}_{\bm{u}}K_{lm}\nabla_jK_k{}^m,\\
	W^{(4)}_4&= \varepsilon^{ijk}K_i{}^l\mathcal{L}_{\bm{u}}K_{jm}\nabla_kK_l{}^m,\\
	W^{(4)}_5&= \varepsilon^{ijk}K_i{}^l\mathcal{L}_{\bm{u}}K_{jm}\nabla^mK_{kl},\\
	W^{(4)}_6&= \varepsilon^{ijk}K_i{}^l\mathcal{L}_{\bm{u}}K_{jm}\nabla_lK_k{}^m.
	\label{eq:W46}
\end{align}

The four families therefore contain $5+2+4+6=17$ basis elements in the $(4,1)$ sector.

\section{Mathematical supplement}
\label{app:math}

For an infinitesimal normal evolution generated by $N u^a$,
with neighboring hypersurfaces identified by this flow, we may
	choose coordinates with vanishing shift vector, so that
$(\partial_t)^a=Nu^a$. In these coordinates,
\begin{equation}
	\delta h_{ij}
	=
	2NK_{ij}\,\delta t,
\end{equation}
which gives
\begin{equation}
	\delta\,{}^{3}\!\Gamma^{k}{}_{ij}
	=
	N C^{k}{}_{ij}\,\delta t. \label{deltaGamma}
\end{equation}
In the above
\begin{equation}
	C^{k}{}_{ij}
	\coloneqq
	\left(\nabla_i+a_i\right)K^{k}{}_{j}
	+\left(\nabla_j+a_j\right)K^{k}{}_{i}
	-\left(\nabla^{k}+a^{k}\right)K_{ij},
	\label{eq:def_C_normal}
\end{equation}
is the normal evolution tensor of the intrinsic connection.
Equivalently,
\begin{equation}
	N C^{k}{}_{ij}
	=
	\nabla_i\left(NK^{k}{}_{j}\right)
	+\nabla_j\left(NK^{k}{}_{i}\right)
	-\nabla^{k}\left(NK_{ij}\right).
	\label{eq:NC_normal}
\end{equation}
Note $C^{k}{}_{ij}$ is a spatial tensor satisfying
$C^{k}{}_{ij}=C^{k}{}_{ji}$.

We first consider the commutators between the Lie derivative along $\bm{u}$ and the intrinsic covariant derivative acting on purely spatial  tensors (with lower indices). These can be derived most simply by working in the zero-shift coordinates, where the Lie derivative along $\bm{u}$  reduces to $(1/N)\partial_t$. The final commutator relations are tensor equations.
For a scalar $S$, one finds
\begin{equation}
	\left[\mathcal{L}_{\bm{u}},\nabla_i\right]S
	=
	a_i\mathcal{L}_{\bm{u}}S.
	\label{eq:comm_scalar}
\end{equation}
For a spatial covector $V_j$,
\begin{equation}
	\left[\mathcal{L}_{\bm{u}},\nabla_i\right]V_j
	=
	a_i\mathcal{L}_{\bm{u}}V_j
	-
	C^{k}{}_{ij}V_k.
	\label{eq:comm_covector}
\end{equation}
For a rank-two covariant spatial tensor $T_{jk}$, one obtains
\begin{equation}
	\left[\mathcal{L}_{\bm{u}},\nabla_i\right]T_{jk}
	=
	a_i\mathcal{L}_{\bm{u}}T_{jk}
	-
	C^{l}{}_{ij}T_{lk}
	-
	C^{l}{}_{ik}T_{jl}.
	\label{eq:comm_tensor}
\end{equation}
More generally, for a purely spatial covariant tensor
$T_{j_1\cdots j_p}$, one has
\begin{equation}
	\left[\mathcal{L}_{\bm{u}},\nabla_i\right]
	T_{j_1\cdots j_p}
	=
	a_i\mathcal{L}_{\bm{u}}T_{j_1\cdots j_p}
	-
	\sum_{r=1}^{p}
	C^{l}{}_{i j_r}
	T_{j_1\cdots j_{r-1}l j_{r+1}\cdots j_p}.
	\label{eq:comm_general}
\end{equation}

We next consider the normal evolution of the intrinsic Ricci tensor
${}^{3}\!R_{ij}$. The Palatini identity for the intrinsic geometry is
\begin{equation}
	\delta\,{}^{3}\!R_{ij}
	=
	\nabla_k\delta\,{}^{3}\!\Gamma^{k}{}_{ij}
	-
	\nabla_j\delta\,{}^{3}\!\Gamma^{k}{}_{ik}.
	\label{eq:palatini_intrinsic}
\end{equation}
Again, we can work in coordinates with vanishing shift. By plugging (\ref{deltaGamma}), the normal evolution of
the intrinsic Ricci tensor can be written as
\begin{align}
	\mathcal{L}_{\bm{u}}{}^{3}\!R_{ij}
	&=
	\frac{1}{N}
	\left[
	\nabla_k\left(NC^{k}{}_{ij}\right)
	-
	\nabla_j\left(NC^{k}{}_{ik}\right)
	\right]
	\nonumber\\
	&=
	\left(\nabla_k+a_k\right)C^{k}{}_{ij}
	-
	\left(\nabla_j+a_j\right)C^{k}{}_{ik}.
	\label{eq:Lie_Ricci_C}
\end{align}
Note \eqref{eq:Lie_Ricci_C} is a tensor equation and is therefore valid in arbitrary coordinates adapted to the same foliation, although it is derived in the zero-shift coordinates.
Using Eq.~\eqref{eq:NC_normal}, this expression may equivalently be written as
\begin{align}
	\mathcal{L}_{\bm{u}}{}^{3}\!R_{ij}
	=
	\frac{1}{N}\Bigl[
	&
	\nabla_k\nabla_i
	\left(NK^{k}{}_{j}\right)
	+
	\nabla_k\nabla_j
	\left(NK^{k}{}_{i}\right)
	\nonumber\\
	&
	-\nabla^{2}\left(NK_{ij}\right)
	-\nabla_i\nabla_j\left(NK\right)
	\Bigr].
	\label{eq:Lie_Ricci_K}
\end{align}


\providecommand{\href}[2]{#2}\begingroup\raggedright\endgroup

\end{document}